\documentclass[12pt]{article}
\usepackage{epsfig}
\usepackage{amssymb,graphicx}
\def\be{\begin{equation}}
\def\ee{\end{equation}}
\def\ba{\begin{array}{c}}
\def\ea{\end{array}}

\def\ben{\[}
\def\een{\]}

\newcommand{\bea}{\begin{eqnarray}}
\newcommand{\eea}{\end{eqnarray}}

\newtheorem{thm}{Theorem}

\newtheorem{lemma}[thm]{Lemma}

\begin{document}

\begin{center}

{\Large \bf {

Polynomial potentials and coupled quantum dots

in two and three dimensions

 }}

\vspace{13mm}

\vspace{3mm}

\begin{center}

\textbf{Miloslav Znojil}


\vspace{4mm}

The Czech Academy of Sciences, Nuclear Physics Institute,

 Hlavn\'{\i} 130,
250 68 \v{R}e\v{z}, Czech Republic\footnote{znojil@ujf.cas.cz},

\vspace{7mm}

Department of Physics, Faculty of Science, University of Hradec
Kr\'{a}lov\'{e},

Rokitansk\'{e}ho 62, 50003 Hradec Kr\'{a}lov\'{e},
 Czech Republic,

\vspace{5mm}

and

\vspace{5mm}

Institute of System Science, Durban University of Technology,

P. O. Box 1334, Durban, 4000, South Africa

\end{center}

\vspace{3mm}

\end{center}

\subsection*{Keywords:}

.

 coupled quantum dots;

 two- and three-dimensional models;

 non-separable polynomial potentials;

 probability density bifurcations;

\subsection*{PACS number:}

PACS

03.65.Ge - Solutions of wave equations: bound states


 \newpage

\subsection*{Abstract}

Polynomial
potentials $V(x)=x^4 + {\cal O}(x^2)$ and
$V(x)=x^6 + {\cal O}(x^4)$
were introduced, in the Thom's
purely geometric classification of bifurcations,
as the benchmark models of the so called
cusp
catastrophes
and of the so called butterfly
catastrophes, respectively.
Due to their
asymptotically confining property,
these two potentials are exceptional,
viz., able to serve similar purposes even
after quantization, in the presence of tunneling.
In this paper the idea is generalized to apply also
to quantum systems in
two and three dimensions.
Two related technical obstacles are addressed,
both connected with the non-separability of
the underlying partial differential Schr\"{o}dinger equations.
The first one [viz., the necessity of a non-numerical localization
of the extremes (i.e., of the minima and maxima) of $V(x,y,\ldots)$]
is resolved via an {\it ad hoc\,} reparametrization of the
coupling constants.
The second one [viz., the necessity of explicit
construction of the low lying bound states $\psi(x,y,\ldots)$]
is circumvented by the restriction of attention to the
dynamical regime in which
the individual minima of $V(x,y,\ldots)$
are well separated, with the potential being locally
approximated by the harmonic oscillator wells
simulating a coupled system of quantum dots
({\it a.k.a.} an artificial molecule).
Subsequently
it is argued that
the measurable characteristics
(and, in particular, the topologically protected
probability-density distributions)
could
bifurcate in specific
evolution scenarios called relocalization catastrophes.

\newpage

\section{Introduction}

The current progress in nanotechnologies leads to multiple
innovations and challenges encountered in the condensed matter
physics and optics as well as in their numerous applications, say,
in material science \cite{wide} or in informatics \cite{daily}. {\it
Pars pro toto} let us mention here, for illustration, the very recent
success in the experimental preparation of isolated tunable quantum
dots and of their interacting multiplets exhibiting, often, various
not quite expected phenomenological features and properties
\cite{triq,bed}.

Initially, the best suited candidates for the theoretical
explanation and description of properties of the quantum dot systems
{\it alias\,} artificial or macroscopic atoms \cite{artiat} and/or
quantum-dot molecules \cite{room} appeared to be the one-, two- or
three-dimensional square wells. Although the typical size of the
systems themselves is given in nanometers, their energy levels still
happened to be discrete and described by Schr\"{o}dinger equation
\cite{qdots}. For theoreticians as well as experimentalists, a
source of certain doubt still lied in the non-analyticity of the
discrete, rectangular square-well potentials $V(x)$, $V(x,y)$ or
$V(x,y,z)$. At the same time, most of the ambitious replacements of
these discontinuous toy-model functions by some more realistic and
smoother shapes led to the bound-state wave functions and energies
given in a purely numerical form.

In recent paper \cite{arpot} we revealed that between the
square-well and the brute-force numerical extremes there may exist a
broad grey zone in which the potentials may be chosen in a specific,
mathematically friendlier polynomial form while the states
themselves can be obtained in a partially non-numerical but still
fairly reliable perturbation-approximation form. In our present
paper we intend to outline an extension of such an idea from the
one-dimensional systems of Ref.~ \cite{arpot} to several
$D-$dimensional scenarios where $D =2$ or $D=3$.

A concise introduction to the field of description of the quantum
dot molecules in $D$ dimensions by analytic potentials will be given
in section \ref{sec2}. We will describe there the path of extension
of the approach and methods of Ref.~\cite{arpot} beyond $D=1$. In
section \ref{tady} we then pick up the first nontrivial polynomials
of order $2N+2=4$ and provide the details of suitable
quantum-dot-molecular models and of the geometry of the related
$D=2$  potentials (in subsection \ref{tadyto}) and of the analogous
$D=3$  potentials (in subsection \ref{tamto}). The next class of
potentials with maximal power $2N+2=6$ will subsequently be studied
in sections \ref{ondyto} and \ref{ondyono}, with the respective
choices of $D=2$ and of $D=3$. Our results may finally be found
discussed and summarized in the last two sections \ref{disco}
and \ref{summary}.

\section{Polynomial potentials in classical and quantum physics\label{sec2}}

\subsection{Partial differential Schr\"{o}dinger equations\label{sec2dot1}}

Phenomenological models of planar or spatial quantum systems are
usually defined via a suitable local potential $V$ and via the
respecctive partial differential Schr\"{o}dinger equation
 \be
 \left[- \frac{\hbar^2}{2\mu_x}\,\frac{d^2}{dx^2}
 - \frac{\hbar^2}{2\mu_y}\,\frac{d^2}{dy^2}+
 V(x,y)
 \right ] \psi_m (x,y) =E_m\,
 \psi_m (x,y)
  \,, \ \ \ \ \
 m=0, 1, \ldots\,
 \label{2dse}
 \ee
or
 \be
 \left[- \frac{\hbar^2}{2\mu_x}\,\frac{d^2}{dx^2}
 - \frac{\hbar^2}{2\mu_y}\,\frac{d^2}{dy^2}
 - \frac{\hbar^2}{2\mu_z}\,\frac{d^2}{dz^2}+
 V(x,y,y)
 \right ] \psi_n (x,y,z) =E_n\,
 \psi_n (x,y,z)
  \,, \ \ \ \ \
 n=0, 1, \ldots\,.
 \label{3dse}
 \ee
A conflict
may then emerge between the needs of
phenomenology (where the realistic potential
may be strongly subject-dependent and far from elementary)
and the costs of the solution.
Those physicists who are
interested in the predictions of the measurable
properties of the spectra of energies (and/or of the other measurable
quantities) are willing to accept any cost, even when a
complicated, brute-force
numerical method had to be used.
In the theory-oriented circles,
the costs tend to be lowered.
In the literature
this
is reflected
by a
disproportionate occurrence of analytic models which are
separable and
solvable in closed form.

We believe that
at least a partial weakening of the conflict
might be provided by the analytic models which only
remain separable and solvable approximately.
In this spirit we will require
a strong form of analyticity in the sense
that the
potentials in Eqs.~(\ref{2dse}) and (\ref{3dse}) (etc)
would be just polynomials (of any even order $2N+2 \geq 4$)
in the $D-$plet of the real spatial coordinates
$x, y, \ldots$. This being said,
we will also require that
 \be
 V(x, y, \ldots) =  (x^2+ y^2+ \ldots)^{N+1}
 + {\cal O}[(x^2+ y^2+ \ldots)^{N}]\,,
 \label{lyacur}
 \ee
i.e., that our partial differential Schr\"{o}dinger equation
still remains
separable at large distances.
For the
sake of definiteness
we will only consider here the first few
simplest even-parity polynomials $V(x, y, \ldots)$
with $2N +2\leq 6$ and with $D\leq 3$.

\subsection{One-dimensional cusp and butterfly catastrophes}

In a broader area of physics the
technical
advantages of working with polynomial potentials
$V(x, y, \ldots)$
emerge not only in
the bound-state theory of quantum mechanics (where the analyticity of
potentials leads to an enormous simplification of perturbation-expansion
techniques of solving Schr\"{o}dinger equations \cite{Kato})
but also in the
qualitative considerations in
classical mechanics and optics.
In the latter setting people usually recall the phenomenology of
bifurcations
in the generic classical dynamical systems where
deep role can be attributed
to general $D=1$ polynomials $V(x)$.
In the form
presented
by Arnold \cite{Arnold} the
mathematical form of
the latter classical-physics idea
inspired also
our recent paper \cite{arpot}. We
revealed and described there certain quantum-mechanical $D=1$ analogues
of the classical bifurcations.
Our present paper can be read as an
explicit, constructive extension
of the $D=1$ results of paper \cite{arpot} to the first two
higher spatial dimensions $D=2$ and $D=3$.

As indicated in review
\cite{Zeeman}, one of the most widely known qualitative classifications of
the parametric
dependence of stationary equilibria and of their bifurcations
in a generic classical dynamical
system is due to
Ren\'{e} Thom \cite{Thom}. In his terminology,
different configurations of possible bifurcations carry
dedicated names of a
``fold catastrophe'', of a ``cusp catastrophe'', etc.
In every one of these arrangements
the eligible equilibria
were identified with
the minima or maxima of certain benchmark polynomials
$V(x, y, \ldots)$,
also known, in this context, as
Lyapunov functions \cite{webcat}.
For example, one of the most popular bifurcation patterns called ``cusp'' has
been assigned, in this scheme, the one-dimensional
quartic-polynomial Lyapunov function
 \be
 V^{[cusp]}(x)=x^4+a\,x^2+b\,x\,.
 \label{lyacu}
 \ee
The superscripted name ``cusp'' refers to the cusped-shaped curve
which separates the two regions supporting one and two stable classical
equilibria (i.e., minima of $V^{[cusp]}(x)$) in the $a-b$ plane of
parameters.

In our paper \cite{arpot}
we pointed out that at least a part
of the classical theory could be perceived as a
semiclassical limit of its quantum-mechanical extension. Every
polynomial Lyapunov function (sampled by Eq.~(\ref{lyacu})) could
then play the role of a potential. In {\it loc. cit.}
we emphasized, in parallel, that besides the limited number of
the candidates taken from the Thom's list, it makes good sense to
extend this list in a way recommended by
Arnold \cite{Arnold}, i.e., to the
infinite family of polynomials of arbitrary degree.

In fact, this degree had to be even because
due to the emergence of quantum tunneling we had to omit all of the
Thom's asymptotically decreasing (i.e., non-confining) benchmark
potentials (like, e.g., $V^{[fold]}(x)=x^3+a\,x$, etc) as unstable.
In fact, it appeared rather unfortunate
that out of the Thom's list of seven basic benchmarks $V(x)$
or $V(x,y)$ we
couldn't accept {\em any\,} Thom's
sample of the $D=2$ Lyapunov function. Even at $D=1$ we
were only left with the cusp of Eq.~(\ref{lyacu}) and
with the only other, so called butterfly Lyapunov function
 \be
 V^{[butter\!fly]}(x)=x^6+a\,x^4+b\,x^3+c\,x^2+d\,x\,.
 \label{buttt}
 \ee
In this way,
we also included the Arnold's higher-degree polynomials. Otherwise,
the number of optional parameters would remain too restricted.

According to the brief subsequent comment \cite{arpot2} the main
weakness of paper \cite{arpot} was that
in spite of its quantum,
genuinely non-geometric and deeply operator-theoretic nature,
its results remained
restricted to the most elementary one-dimensional class of toy models. In
what follows, we intend to fill the gap. We will
consider just the cusp and butterfly quantum potentials
$V(x)$, but we will
extend the results
of Ref.~\cite{arpot} and cover both the two- and
three-dimensional generalized scenarios.

\subsection{Asymptotically separable
polynomial potentials}

In conventional Schr\"{o}dinger equations the mass-dependence constants
$\Lambda_j={\hbar^2}/{2\mu_{j}}$
may be treated as a
$D-$plet of variable phenomenological parameters.
This would open the way to the study of the
semiclassical limit where these constants are small,
$\Lambda_j \approx 0$. Under this assumption one reveals
that the low-lying energy levels $E_0$, $E_1$, \ldots
converge to the same value which is equal to the absolute
minimum $V(X,Y,\ldots)$ of the potential at $x=X$, $y=Y$, etc.
In this
sense the semiclassical limit mediates a connection between
quantum mechanics and, say, the Thom's \cite{Thom} classical theory
describing equilibria and the bifurcations {\it alias\,}
catastrophes in non-quantum dynamical systems (see also the thorough
dedicated reviews \cite{revcat}).

The latter observation motivated our
quantum-mechanical $D=1$ study \cite{arpot}. Potentials were
chosen there, for the sake of simplicity, in the form of the
Arnold's \cite{Arnold}
even-parity
polynomials $V(x)= x^{2N+2}+ {\cal
O}(x^{2N})$ of any degree $2N+2$. We pointed out that in spite
of the emergence of tunneling at non-vanishing $\Lambda_j \neq 0$,
the Thom's classical catastrophes characterized by an abrupt change
of the observable features of the system after a minor change of the
parameters can find certain analogues in the genuine quantum
context.

In the opposite, ultra-quantum limit with large $\Lambda_{j} \to
\infty$ the distance between the minimum $V(X,Y,\ldots)$ and the
ground-state energy $E_0$ (as well as between $E_0$ and $E_1$, etc)
happens to grow. The shape of the potential
remains relevant even far from the origin.
This observation forces one to
eliminate the odd-degree polynomials (like $V^{[fold]}(x)=x^3
+ ax$, etc) from the Thom's list as unstable and unacceptable
after quantization.

We will only consider
the reality-reflecting
dimensions $D=2$ and $D=3$ in what follows. Moreover,
we will simplify our task by admitting just
Schr\"{o}dinger equations which are asymptotically separable.
This will be achieved via
constraint (\ref{lyacur}).
Such an additional
simplification
is purely formal because we will not use the
asymptotic
separability of Schr\"{o}dinger equations,
treating constraint~(\ref{lyacur}) merely as a
means of elimination of qualitatively less
essential coupling constants.

The dynamics-modifying role of at least some of these
omitted constants may
be transferred to the mass terms $\Lambda_j$.
Having clarified their direct role of mediators of the classical and/or
ultra-quantum limits, we shall skip this analysis, and we shall
keep all of the mass terms $\Lambda_j$
equal to one in what follows.

\section{Cusp potentials in more than one dimension\label{tady}}

Paper \cite{gilmore0} should bbe cited here as offering one of the
pioneering studies of
the one-dimensional Schr\"{o}dinger equation
in which the role of the potential is played by the
elementary $D=1$ version of
the cusp Lyapunov function.
The authors performed, in particular,
a detailed analysis of the
localization and delocalization
of the probability distributions.
The paper strongly motivated also our present analysis of the
possibilities of a qualitative understanding of the
$D=2$ and $D=3$ Schr\"{o}dinger equations containing the
generalized potentials of the cusp type.

The apparently purely numerical,
non-separable partial-differential nature
of such a schematic bound state project
redirected, presumably, attention of
may interested researchers
to the study of the full-fledged, realistic quantum systems in which the
cusp-catastrophic phenomena only re-emerged as a consequence
of their constructive analysis. For
an illustration of this trend let us only recall
the very recent study \cite{gilmore3} in which the term
``quantum catastrophe'' has been attributed to
an open quantum system
(sampled by a Josephson junction made
of two Bose-Einstein condensates) in which
the authors apply the Thom's concepts of the
fold and cusp catastrophes to the caustics
in the number-difference probability distribution.

In this context our present study of the $D=2$ and $D=3$ models
looks, and is, rather naive. One has to emphasize that
the above-mentioned direction of considerations
(cf. also, in this respect, \cite{gilmore1,gilmore2})
really represents just one of many innovative
applications of the classical Thom's theory.

\subsection{Two dimensions\label{tadyto}}

\begin{figure}[h]                    
\begin{center}                         
\epsfig{file=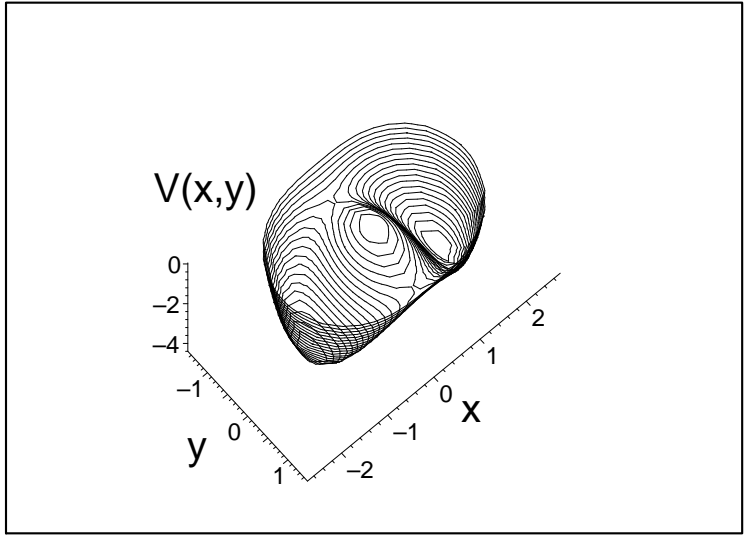,angle=270,width=0.56\textwidth}
\end{center}    
\vspace{2mm} \caption{$V(x,y)\leq 0$ part of potential (\ref{[4]})
with the two equal minima $V(\pm \alpha,0)=-\alpha^4$ and with the two equal
saddle-point values $V(\pm \beta,0)=-\beta^4$
at parameters $\alpha=1.4$
and $\beta=1$.
 \label{fione}
 }
\end{figure}

Two-parametric cusp-type, asymptotically separable potential
 \be
 V^{[cusp]}(x,y)=r^4-2\,\alpha^2\,x^2-2\,\beta^2\,y^2\,,
 \ \ \ \ \ r^2=r^2(x,y)=x^2+y^2\,
 \label{[4]}
 \ee
with, say, $\alpha^2>\beta^2$ has its real extremes (or, more
precisely, the points of stationarity -- if any) at the roots
$(X,Y)$ of the following coupled pair of polynomial equations
 \be
 \partial_X V^{[cusp]}(X,Y)=4\,X\,(R^2-\alpha^2)=0\,,
 \ \ \ \ \
 \partial_Y V^{[cusp]}(X,Y)=4\,Y\,(R^2-\beta^2)=0\,
 \ee
where we abbreviated $R^2=R^2(X,Y)=X^2+Y^2$. On this background it
is now rather lengthy but entirely straightforward to reveal and
prove that the local points of stationarity occur at $(X,Y)=(0,0)$
(this is a local or absolute maximum at the origin), at $(X,Y)=(0,\pm \beta)$
(two local saddle points located off the origin), and at $(X,Y)=(\pm
\alpha,0)$ (the other two local or absolute minima off the origin -- see
Fig.~\ref{fione} for an illustrative example).



Near both of the non-central local or absolute minima
we may re-arrange our polynomial potential yielding
the exact, terminating Taylor series,
 $$
 V^{[cusp]}(x,y)=
 -{{\it \alpha}}^{4}+4\,{{\it \alpha}}^{2} \left( x \mp {\it \alpha}
 \right) ^{2}+2\,
 \left( {{\it \alpha}}^{2}-{{\it \beta}}^{2} \right) {y}^{2}\pm
  4\,{\it \alpha}\, \left( x \mp {\it \alpha} \right) ^{3}\pm
 4\,{\it \alpha}\,{y}^{2}
\left( x \mp { \it \alpha} \right) +
 $$
 $$
  + \left( x \mp {\it \alpha} \right)^{4}+2\,{y}^{2}
  \left( x \mp { \it \alpha} \right) ^{2}+{y}^{4}\,.
 $$
As long as we assumed that $\alpha^2>\beta^2$ we may immediately conclude that
these extremes in fact represent the two absolute minima of the potential.
We may also notice that
during the tentative decrease of $\alpha^2$ to its boundary value
$\beta^2$ (i.e., during a transition to the
rotationally
invariant and exactly separable, continuous-spectrum-supporting
Mexican-hat limit of the potential) the leading-order
approximation of the low lying discrete spectrum generated by the
locally dominant harmonic-oscillator
well loses its precision because the well itself becomes too much
protruded and less and less confining along the $y-$axis.

\subsection{Three dimensions\label{tamto}}

At $D=3$ the positions $(X,Y,Z)$ of the real points of stationarity
of the asymptotically separable quartic potential
 \be
 V^{[cusp]}(x,y,z)=r^4-2\,\alpha^2\,x^2
 -2\,\beta^2\,y^2-2\,\gamma^2\,z^2\,,
 \ \ \ \ \ r^2=r^2(x,y)=x^2+y^2+z^2\,
 \label{[4d3]}
 \ee
will vary with the triplet of parameters such that, say,
$\alpha^2>\beta^2>\gamma^2$. These positions
will be determined
by the three coupled polynomial equations
 \ben
 \partial_X V^{[cusp]}(X,Y,Z)=4\,X\,(R^2-\alpha^2)=0\,,
 \een
 \ben
 \partial_Y V^{[cusp]}(X,Y,Z)=4\,Y\,(R^2-\beta^2)=0\,,
 \een
 \ben
 \partial_Z V^{[cusp]}(X,Y,Z)=4\,Z\,(R^2-\gamma^2)=0\,
 \een
where we abbreviated $X^2+Y^2+Z^2=R^2(X,Y,Z)=R^2$.

\begin{table}[h]
\caption{Points of stationarity of potential (\ref{[4d3]})}
 \label{dowe} \vspace{.1cm}
\centering
\begin{tabular}{||c|c|c|c||}
\hline
  \hline
  &  {\rm (X,Y,Z)} &
   &  V(X,Y,Z)
         \\
 \hline
      \hline
  1& {\rm  {\rm  {\rm  (0,0,0) }}}
   &  {\rm  (central\ local\ maximum) }& 0\\
  2&  (0,0,$\pm \gamma$)
   &  {\rm  (two saddle points) }
   &  $-\gamma^4$\\
  3& {\rm  {\rm  {\rm  (0,$\pm \beta$,0) }}}
   &  {\rm  (two saddle points) }
   &  $-\beta^4$\\
  4& {\rm  {\rm  {\rm  ($\pm \alpha$,0,0) }}}
   &  {\rm  (two absolute minima) }
   &  $-\alpha^4$\\
 \hline
 \hline
\end{tabular}
\end{table}

In a small vicinity of every stationarity point the
potential may be replaced again by its respective leading-order
approximation. Thus, near the origin, i.e., near the local maximum at
$(X,Y,Z)=(0,0,0)$ we reveal, in its small vicinity, an inverted-cup shape,
 \be
 V(x,y,z) \approx -2\,{{\it {\alpha}}}^{2}{x}^{2}
 -2\,{{\it {\beta}}}^{2}{y}^{2}-2\,{{\it {\gamma}}}^{2}{z}^{2}\,.
 \ee
Near another stationary point $(X,Y,Z)=(\pm \alpha,0,0)$) we discover
a pair of absolute minima,
 \be
 V(x,y,z) \approx -{{\it {\alpha}}}^{4}
 +4\,{{\it {\alpha}}}^{2} \left( x \mp
 {\it {\alpha}} \right) ^{2}+
 \left( -2\,{{\it {\beta}}}^{2}
 +2\,{{\it {\alpha}}}^{2} \right) {y}^{2}+ \left( -
 2\,{{\it {\gamma}}}^{2}+2\,{{\it {\alpha}}}^{2} \right) {z}^{2}\,.
 \ee
In this
approximation, due to the emergent
separability and exact solvability of the
reduced model,
the numerical construction of
the low-lying bound states may be replaced by
the leading-order
closed formulae followed, if needed, by the
perturbation-theory evaluation of corrections
(these exercises are left to the reader).

At the other two stationary-point
candidates  $(X,Y,Z)=(0,\pm \beta,0)$) and
$(X,Y,Z)=(0,0,\pm \gamma)$ we obtain the respective
local approximations
 \be
 V(x,y,z) \approx
 -{{\it {\beta}}}^{4}+ \left( -2\,{{\it {\alpha}}}^{2}
 +2\,{{\it {\beta}}}^{2} \right) {
 x}^{2}
 +4\,{{\it {\beta}}}^{2} \left( y \mp  {\it {\beta}} \right) ^{2}+ \left(
 2\,{{ \it {\beta}}}^{2}-2\,{{\it {\gamma}}}^{2} \right) {z}^{2}
  \ee
and
 \be
 V(x,y,z) \approx
 -{{\it {\gamma}}}^{4}+ \left( 2\,{{\it {\gamma}}}^{2}
 -2\,{{\it {\alpha}}}^{2} \right) {x
 }^{2}+ \left( 2\,{{\it {\gamma}}}^{2}-2\,{{\it {\beta}}}^{2} \right)
 {y}^{2}
 +4\,{ {\it {\gamma}}}^{2} \left( z \mp {\it {\gamma}} \right)^{2}\,,
  \ee
i.e., the saddle-point behavior.
The list is complete.
The situation is summarized
in Table \ref{dowe}.

One could also characterize the
global
shapes of the potentials by displaying the pictures of
their separate $x=0$ or $y=0$ or $z=0$ sections. Clearly, the
simplicity of the model makes such an illustration redundant. Moreover,
the resulting pictures would all look precisely like
Fig.~\ref{fione} above.


%

\section{Butterfly potential in two dimensions \label{ondyto}}

In \cite{arpot} we
considered the first nontrivial confining $D=1$
potential (\ref{buttt}) in a spatially symmetric
special case with $b=d=0$. We choose negative
$a=-3(\alpha^2+\beta^2)$ together with positive
$c=3\alpha^2\gamma^2$ where
$\gamma^2=\gamma^2(\alpha,\beta)=\alpha^2+2\,\beta^2$. This enabled
us to guarantee that our one-dimensional potential
$V^{[butter\!fly]}(x)$ had a triple-well shape admitting the three
well-separated local minima (mimicking quantum dots)
such that $V(x)=0$ vanished at $x=0$ and such
that $V(\pm \gamma)=(\alpha^2-\beta^2)\gamma^4$ could have both signs.
Now we intend to introduce an analogue of this model at $D=2$.

\subsection{Basic features}

In the context of quantum physics
we are interested in the identification of the
situation in which a small change of
parameter would
cause
a ``relocalization catastrophe'', i.e., an abrupt jump from
the single-centered measurable probability density localized near
the origin to the topologically different quantum state with the
two-centered probability density localized far from the origin.

At $D=2$, having the same quantum-dot-related
motivation as at $D=1$,
the
five-parametric two-dimensional sextic-polynomial potential
 \be
 V^{[butter\!fly]}(x,y)=r^6-3\,a\,x^4-3\,u\,x^2y^2
 -3\,b\,y^4+3\,c\,x^2+3\,d\,y^2\,,
 \ \ \ \ \ r^2=r^2(x,y)=x^2+y^2\,
 \label{[6]}
 \ee
will be considered with
positive $a$, $b$, $c$ and $d$, and with a real $u$.

First of
all we will have to find the real extremes of the potential if any.
Along the same lines as above the coordinates $(X,Y)$ of these
extremes must satisfy the following two coupled polynomial equations
 \be
 \partial_X V^{[butter\!fly]}(X,Y)=
 6X\,(R^4-2\,a\,X^2-u\,Y^2+c)
 =0\,,
 \ee
 \be
 \partial_Y V^{[butter\!fly]}(X,Y)
 =
 6Y\,(R^4-u\,X^2-2\,b\,Y^2+d)
 =0\,.
 \ee
The trivial central extreme at $(X,Y)=(0,0)$ is a local minimum
because whenever $x$ and $y$ remain sufficiently small we can simply
reorder and truncate,
 \be
 V^{[butter\!fly]}(x,y)=3\,c\,x^2+3\,d\,y^2
 +3\,a\,x^4+3\,u\,x^2y^2
 +3\,b\,y^4+r^6 \approx 3\,c\,x^2+3\,d\,y^2\,.
 \label{[6b]}
 \ee
We easily identify the four off-central extremes with
$(X,Y)=(0,\pm Y_\pm)$ and
 \be
 Y_\pm^2 =b\pm \sqrt{b^2-d\ }
  \ee
plus, similarly, the other four extremes with $(X,Y)=(\pm X_\pm,0)$
where
 \be
 X_\pm^2 =a\pm \sqrt{a^2-c\ }\,.
  \ee
In the light of the latter formulae it makes sense to reparametrize
the couplings,
 \be
 a=\alpha_x^2+\beta_x^2\,,
 \ \ \ \
 c=\alpha_x^2(\alpha_x^2+2\,\beta_x^2)=\alpha_x^2\gamma_x^2\,,
 \ee
 \be
 b=\alpha_y^2+\beta_y^2\,,
 \ \ \ \
 d=\alpha_y^2(\alpha_y^2+2\,\beta_y^2)=\alpha_y^2\gamma_y^2\,.
 \ee
A complete parallel emerges with the one-dimensional case. The
reparametrization yields the following, most
elementary parameter-interpretation formulae
 \be
 X_-^2 =\alpha_x^2
 \ \ < \ \
 X_+^2 =\gamma_x^2\,
 \ee
and
 \be
 Y_-^2 =\alpha_y^2
 \ \ < \ \
 Y_+^2 =\gamma_y^2\,.
 \ee
Considering just the real points of stationarity (having an
elementary geometric interpretation and being localized along the
coordinate axes) one reveals a one-to-one correspondence between the
guarantees of their reality at $D=2$ and at $D=1$.


\begin{figure}[h]                    
\begin{center}                         
\epsfig{file=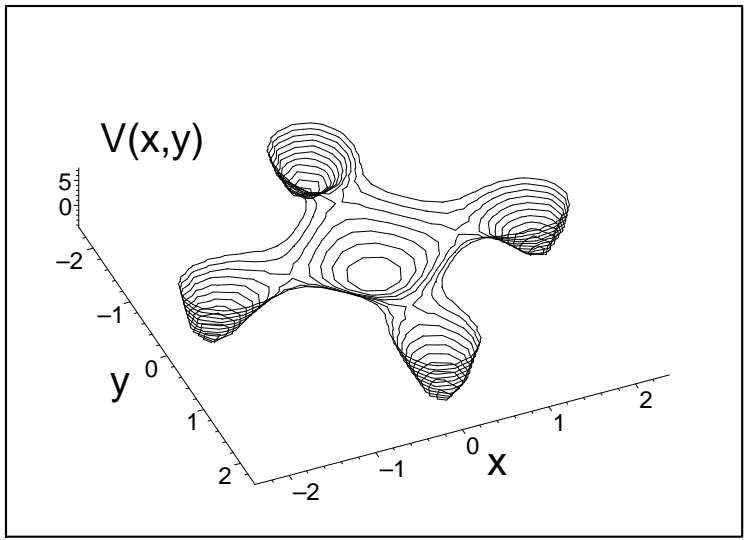,angle=270,width=0.56\textwidth}
\end{center}    
\vspace{2mm} \caption{The low-lying part of potential
(\ref{[6]}) in the $(x-y)$-symmetric
five-deep-dots regime,
with $V(x,y)<7.5$, $\alpha_{x,y}=1$, $\gamma_{x,y}=1.9$ and with $u=-16/3$.
 \label{fi2}
 }
\end{figure}

\subsection{Points of stationarity off the coordinate axes\label{tadyhle}}

What remains to be discussed is the localization of the $D=2$ points
of stationarity  which do not lie on one of the coordinate axes.
Under the assumption $X \neq 0 \neq Y$ these points must be sought
as the roots of the set of equations
 \be
  R^4-u\,R^2+c
 =(2\,a-u)\,X^2\,,
 \ \ \ \ \
 R^4-u\,R^2+d
 =(2\,b-u)\,Y^2\,.
 \ee
These two relations may be reinterpreted as the respective
definitions of $X^2$ and $Y^2$ in terms of an unknown quantity
$R^2$. The condition of compatibility $X^2+Y^2=R^2$ is to be added
at the end. This yields the exactly solvable quadratic equation for
$R^2$ leading to the two closed-form roots
 \be
 R^2=R^2_\pm(u)=\frac{uz(u)+1\pm \sqrt{[uz(u)+1]^2-4z(u)w(u)\ }}{2z(u)}\,
 \label{fofo}
 \ee
where we abbreviated
 \be
 w(u)=\frac{c}{2a-u} + \frac{d}{2b-u} \,,
 \ \ \ \
 z(u)=\frac{1}{2a-u} + \frac{1}{2b-u} \,.
 \ee
It is rather straightforward to
sample the variability of descriptive
features of
our asymptotically circular (i.e., asymptotically separable)
model.
At the positive and large values of $u$, indeed, both
of our above-mentioned auxiliary functions $ w(u)$ and $ z(u)$ are
small and negative. In such a dynamical regime the parameters
$R_\pm(u)$ defined by formula (\ref{fofo}) cannot be real so that we
only have to consider the real stationary points such that $X=0$ or
$Y=0$ or $X=Y=0$. The corresponding shape of the potential is then
sampled in Figure \ref{fi2}. We choose there the values of
parameters $\alpha_x=\alpha_y=1$, $\gamma_x=\gamma_y=1.9$ and a
fairly large $|u|$. In such an arrangement the potential
still has the standard and expected single local minimum in the
origin. This is accompanied by the four absolute minima which are
perceivably deeper and mutually well separated.
We may conclude that after transition from one to two dimensions the
shape of the butterfly-inspired sextic polynomial potentials becomes
richer and more flexible than a priori expected.

An exhaustive qualitative
classification and discussion of our fully general five-parametric
model would be too long. Even any
non-exhaustive, illustrative description of the details of the
parametric dependence of the bound states
lies beyond the scope of our present study.
Still, its fairly detailed outline
considering a few most elementary special cases of
potential (\ref{[6]})
containing just three independent parameters
and possessing a
triplet of minima separated by pronounced barriers
is under preparation at present \cite{arpot2}.

\section{Butterfly potential in three dimensions\label{ondyono}}

\subsection{Points of stationarity on coordinate axes}

In three dimensions let us abbreviate $r^2=r^2(x,y,z)=x^2+y^2+z^2$
and consider the following nine-parametric family of potentials
 \ben
 V^{[butter\!fly]}(x,y,z)=r^6-3\,a\,x^4
 -3\,b\,y^4
 -3\,c\,z^4
 -3\,u\,x^2y^2
 -3\,v\,x^2z^2
 -3\,w\,y^2z^2+
 \een
 \be
 +3\,p\,x^2+3\,q\,y^2+3\,s\,z^2\,.
 \label{[6d3]}
 \ee
We assume that $p$, $q$ and $s$ are positive so that we
recognize, immediately, the existence of a local minimum at
$(x,y,z)=(X,Y,Z)=(0,0,0)$. Indeed, near this point we have, in the
leading-order approximation,
 \be
 V^{[butter\!fly]}(x,y,z)\approx 3\,p\,x^2+3\,q\,y^2+3\,s\,z^2
 \,.
 \label{[6bd3]}
 \ee
The search for the other minima, maxima or saddle points of our
sixth-degree polynomial of three variables (\ref{[6d3]})
may proceed in
a way paralleling our preceding
constructions. First, we differentiate
 \ben
 \partial_X V^{[butter\!fly]}(X,Y,Z)=
 6X\,(R^4-2\,a\,X^2-u\,Y^2-v\,Z^2+p)
 \,,
 \een
 \ben
 \partial_Y V^{[butter\!fly]}(X,Y,Z)
 =
 6Y\,(R^4-2\,b\,Y^2-u\,X^2-w\,Z^2
 +q)
 \,,
 \een
 \ben
 \partial_Z V^{[butter\!fly]}(X,Y,Z)
 =
 6Z\,(R^4-2\,c\,Z^2-v\,X^2-w\,Y^2
 +s)
 \,
 \een
and reparametrize the couplings,
 \ben
 a=\alpha_x^2+\beta_x^2\,,
 \ \ \ \
 p=\alpha_x^2(\alpha_x^2+2\,\beta_x^2)=\alpha_x^2\gamma_x^2\,,
 \een
 \ben
 b=\alpha_y^2+\beta_y^2\,,
 \ \ \ \
 q=\alpha_y^2(\alpha_y^2+2\,\beta_y^2)=\alpha_y^2\gamma_y^2\,,
 \een
 \ben
 c=\alpha_z^2+\beta_z^2\,,
 \ \ \ \
 s=\alpha_z^2(\alpha_z^2+2\,\beta_z^2)=\alpha_z^2\gamma_z^2\,.
 \een
This identifies those stationary
points which lie on one of the coordinate axes. Their location and
some properties are sampled in Table~\ref{rydo}.


\begin{table}[h]
 \caption{Points of
 stationarity of potential (\ref{[6d3]}) along coordinate axes.
 Parameters
 and their differences are assumed partially ordered, with
 $\gamma_x^2>\gamma_y^2>\gamma_z^2$ and
 $\alpha_x^2-\beta_x^2<\alpha_y^2-\beta_y^2<\alpha_z^2-\beta_z^2<0$.
 }
 \label{rydo} \vspace{.1cm}
\centering
\begin{tabular}{||c|c|c|c||}
\hline
  \hline
  &  {\rm (X,Y,Z)} &
   &  V(X,Y,Z)
         \\
 \hline
      \hline
  1& {\rm  {\rm  {\rm  (0,0,0) }}}
   &  {\rm  (single central local minimum) }& 0\\
  2&  (0,0,$\pm \alpha_z$)
   &  {\rm  (two saddle points) }
   &  $\alpha_z^4(\alpha_z^2+3\,\beta_z^2)$\\
 3&  (0,$\pm \alpha_y$,0)
   &  {\rm  (two saddle points) }
   &  $\alpha_y^4(\alpha_y^2+3\,\beta_y^2)$\\
 4&  ($\pm \alpha_x$,0,0)
   &  {\rm  (two saddle points) }
   &  $\alpha_x^4(\alpha_x^2+3\,\beta_x^2)$\\
 5&  (0,0,$\pm \gamma_z$)
   &  {\rm  (saddle points or local minima) }
   &  $(\alpha_z^2-\beta_z^2)\gamma_z^4$\\
 6&  (0,$\pm \gamma_y$,0)
   &  {\rm  (saddle points or local minima) }
   &  $(\alpha_y^2-\beta_y^2)\gamma_y^4$\\
 7&  ($\pm \gamma_x$,0,0)
   &  {\rm  (saddle points or absolute minima) }
   &  $(\alpha_x^2-\beta_x^2)\gamma_x^4$\\
 \hline
 \hline
\end{tabular}
\end{table}

\subsection{Points of stationarity off the coordinate axes}

Off the coordinate axes
the family of points of stationarity $(X,Y,Z)$
may be separated
into the three perpendicular-plane subfamilies (such that $X=0$
but $Y \neq 0 \neq Z$, etc) and the fourth, bulk space subfamily
with all of the stationary-point coordinates
$X$, $Y$ and $Z$ non-vanishing.
In the former three cases the
answers represented by the closed-form roots of a quadratic
algebraic equation
remain
fully analogous to their $D=2$ predecessors
(in fact, the $Z=0$ root is given
precisely by Eq.~(\ref{fofo}) above).
In the last, bulk-space case the real triplet of quantities
$(X,Y,Z)$ (if any) must coincide with the roots of the reduced
coupled algebraic equations
 \be
  R^4-2\,a\,X^2-u\,Y^2-v\,Z^2+p=0
 \,,
 \label{11}
 \ee
 \be
 R^4-2\,b\,Y^2-u\,X^2-w\,Z^2
 +q=0
 \,,
 \ee
 \be
  R^4-2\,c\,Z^2-v\,X^2-w\,Y^2
 +s=0
 \,.
 \label{33}
 \ee
After transition from $D=2$ to $D=3$
the process of the solution of these equations does not change too much.
In a preparatory step
we may again temporarily
treat
the symbol $R^4$ as an arbitrarily variable
real parameter. Relations (\ref{11}) --
(\ref{33}) may be then re-read as the linear algebraic set
 \be
 \left[ \begin {array}{ccc} 2\,{\it a}&u&v
 \\\noalign{\medskip}u&2\,{\it b}&w
 \\\noalign{\medskip}v&w&2\,{\it c}\end {array} \right]\,
 \left[ \begin {array}{c}
 X^2
  \\\noalign{\medskip}Y^2
   \\\noalign{\medskip}Z^2
 \end {array} \right]=\left[ \begin {array}{c}
 R^4+p
  \\\noalign{\medskip}R^4+q
   \\\noalign{\medskip}R^4+s
 \end {array} \right] \,.
 \label{tarov}
 \ee
These equations
are solvable via an elementary matrix inversion
determining
quantities $X^2$, $Y^2$ and $Z^2$ as linear functions of our auxiliary
parameter $R^4$.
The insertion of the resulting
definitions $X^2=X^2(R^4)$,  $Y^2=Y^2(R^4)$ and  $Z^2=Z^2(R^4)$
in the constraint
$R^2=X^2+Y^2+Z^2$ finally yields quadratic equation
 \be
 R^2=X^2(R^4)+Y^2(R^4)+Z^2(R^4)
 \label{jieq}
 \ee
for unknown $R^2$ as well as the two
explicit forms $R^2_\pm$ of
its root.

\section{Discussion\label{disco}}

Even in the latter,
butterfly model at $D=3$,
the exhaustive analysis of the
domains of parameters in which the roots $R$
are real and positive would be lengthy but still
straightforward.
In particular, the shape of the potential will stay
characterized,
in the most interesting scenarios, by the presence of several
pronounced local minima
of the depths under our control.
They would be
mutually separated by barriers
so that the low-lying states would be localized near absolute
minimum or minima.

Naturally, all of these observations would be supported by
the corresponding closed formulae. At the same time,
one has to pay main attention to the
special cases in which these formulae simplify.
Thus, in a way
promoted in our methodical guide \cite{arpot},
one should pay particular attention
to the
most user-friendly
cases in which the parameters are large,
$\alpha_{x,y,z}^2 \gg 1$ as well as $\beta_{x,y,z}^2 \gg 1$.
Indeed, precisely in this dynamical regime
our potentials happen to
exhibit the phenomenologically most interesting
shapes as
sampled in Fig.~\ref{fi2}. Let us now describe and discuss
a few samples
of these simplified scenarios in more detail.

\subsection{The case of small $u$, $v$ and $w$}

The
simplifications of mathematics caused by the
use of the large parameters $\alpha$ and $\beta$
can be sought along the lines outlined
in section \ref{tadyto} above.
New complications arise due to the three
parameters $u$, $v$ and $w$
which need not be positive.
Fortunately,
the inversion in Eq.~(\ref{tarov}) becomes trivial whenever
their size is negligible.
The explicit simplified form of our
quadratic Eq.~(\ref{jieq}) then reads
 \be
 (ab+ac+bc)\,R^4-2abc\,R^2+sab+qac+pbc
 =0\,.
 \ee
In the isotropic limit
$\alpha_x=\alpha_y=\alpha_z=\alpha$ and
$\beta_x=\beta_y=\beta_z=\beta$,
for example,
we get the roots
 \be
 R^2_\pm = \frac{1}{3}
 \,
 \left (\alpha^2+\beta^2 \pm \sqrt{
 \beta^4-8\,\alpha^2\,(\alpha^2+2\beta^2)
 }
 \right )\,.
 \label{rooib}
 \ee
We arrive at the following criterion
guaranteeing the reality (i.e., the existence) of the
points of stationarity.

\begin{lemma}
Both roots (\ref{rooib}) are (simultaneously)
real and positive if and only if
the ratio $\xi=\alpha/\beta$
is  small
enough,
such that $ \xi^2\,(2+\xi^2)\leq 1/8$, i.e., such that
$\xi \leq 1/2\,\sqrt {-4+3\,\sqrt {2}} \approx
0.2462928572$.
\end{lemma}

\subsection{The case of large $u$, $v$ or $w$}

In the opposite extreme with a
scale parameter $\lambda \gg 1$ such that
$\alpha_{x,y,z} = {\cal O}(\lambda)$ and
$\beta_{x,y,z} = {\cal O}(\lambda)$
while
$|u| = {\cal O}(\lambda^2)$
or $|v| = {\cal O}(\lambda^2)$ or $|w| = {\cal O}(\lambda^2)$
we may omit the other,
newly subdominant terms from Eq.~(\ref{tarov})
and obtain its simplified version
 \be
 \left[ \begin {array}{ccc} 0&u&v
 \\\noalign{\medskip}u&0&w
 \\\noalign{\medskip}v&w&0\end {array} \right]\,
 \left[ \begin {array}{c}
 X^2
  \\\noalign{\medskip}Y^2
   \\\noalign{\medskip}Z^2
 \end {array} \right]=\left[ \begin {array}{c}
 R^4+p
  \\\noalign{\medskip}R^4+q
   \\\noalign{\medskip}R^4+s
 \end {array} \right] \,
  \label{tarovni}
 \ee
with the $R^4-$parametrized solution
 \be
  \left[ \begin {array}{c}
 X^2(R^4)
  \\\noalign{\medskip}Y^2(R^4)
   \\\noalign{\medskip}Z^2(R^4)
 \end {array} \right]=\frac{1}{2uvw}\,
 \left[ \begin {array}{ccc} -{w}^{2}&vw&uw
 \\\noalign{\medskip}vw&-{v}^{2}&vu
 \\\noalign{\medskip}uw&vu&-{u}^{2}\end {array} \right]
  \,
 \left[ \begin {array}{c}
 R^4+p
  \\\noalign{\medskip}R^4+q
   \\\noalign{\medskip}R^4+s
 \end {array} \right] \,.
  \label{tarovni}
 \ee
Reconstruction rule (\ref{jieq}) then
leads to quadratic equation
 \ben
 (-{w}^{2}+2\,vw+2\,uw-{v}^{2}+2\,vu-{u}^{2})\,R^4
 -2uvw\,R^2-
 \een
 \be
  -p{w}^{2}+qvw+suw+pvw-{v}^{2}q+svu+puw+qvu-s{u}^{2}=0\,.
 \ee
Formula for its roots $R^2_\pm$ is elementary but
lengthy to print.
Fortunately, in the isotropic limit with $u=v=w$
it remains compact and transparent,
 \be
 R^2_\pm=\frac{1}{3}\,u \pm \sqrt{u^2-3p-3q-3s\ }\,.
 \label{boi}
 \ee
As long as the values of
$p$, $q$ and $s$ are assumed large and positive,
we arrive at another elementary criterion
guaranteeing the existence of the
points of stationarity lying off the coordinate axes.

\begin{lemma}
Solution (in fact, both roots) (\ref{boi})
remains real and positive (i.e., acceptable) if
and only if the value of $u$ is positive and,
moreover, sufficiently large,
 \be
 u=v=w \geq \sqrt{3\,(p+q+s)\ }= {\cal O}(\lambda^2)\,.
 \ee
\end{lemma}

\subsection{Experimental aspects}

A key descriptive feature of the present $D=2$ and $D=3$ polynomial
potentials $V(x,y,\ldots)$ is a nontrivial though still constructive
and controllable variability of the dominance of the subsets of
their local minima. This implies that at a fixed set of parameters
the bound state densities reflecting the dominance form a
topologically protected feature of the system. At the same time, we
saw that along certain parameter-subdomian boundaries
this protection may be made fragile via a comparatively
very small change of the parameters.

This is one of the not entirely obvious manifestations of certain
incomplete qualitative parallels between the Thom's classical
classification of catastrophes where there is, by definition, no
tunneling, and its present quantum-theoretical analogue in which the
tunneling is, naturally, present and allowed.
In fact, the control of the intensity of the tunneling is only one
of the aspects of the possible measurable features of the systems
which would be sufficiently well described by one of the present
polynomial potentials. The
phenomenon may be suppressed via a heightening and/or broadening of
the barriers. Along these lines one could really expect an
enhancement of our understanding of ``the confinement of 2D
electrons in customizable potentials'' \cite{bed}, etc.

Another, more specific experimental benefit may be seen in the
smoothness of the shapes of the polynomial potentials which we
sampled above. Indeed, such potentials might fit the realistic
shapes of quantum dots much better than the discontinuous square
wells. After all, the sets of quantum dots are quite difficult to
prepare. Typically, quoting again the words of Ref.~\cite{bed}, this
has been achieved by ``using a low-temperature scanning tunneling
microscope'' as a microscopic drilling tool which may ``create and
directly image a new type of coupled quantum dot system in
graphene''.

Even in our present, most elementary polynomial-potential models we
revealed a fairly high sensitivity of their shapes and spectra to the
parameters. For this reason it seem really challenging to search for
the abrupt probability-redistribution transitions called
``relocalization catastrophes'', with a perceivable phenomenological
appeal. In a longer perspective the latter transitions may be
expected to be exploited, e.g., in the quantum versions of the
information processing.

Once we restrict attention again just to our
most complicated $D=3$ polynomial
potential (\ref{[6d3]}) we may really confirm that whenever some of
the relevant parameters become large (forming a regime of particular
interest), the explicit determination of the positions of all of
the local minima of $V(x,y,z)$ happened to be fairly user-friendly,
i.e., as straightforward as at $D=2$. We saw that the analogy with
the $D=2$ predecessor survived also in the vicinity of all of the
local extremes, minima or maxima. With good precision, the shape of
the potential remained approximately equal there to a separable,
exactly solvable harmonic-oscillator well.
As a consequence of such an apparently purely technical
observation, a non-numerical, closed-formula access has been open
not only to the low-lying energies $E_0$, $E_1$, \ldots but also,
more importantly, to the dominant components of the density
distributions $|\psi_0|$, $|\psi_1|$, \ldots\,. Indeed, as long as
one knows the parameter-dependence of the latter low-lying
harmonic-oscillator states {\em locally}, i.e., near {\em every\,}
(i.e., not necessarily just absolute)  minimum, one can determine
the critical values of parameters at which the dominance gets
transferred from an ``old'' absolute minimum or minima to the
``new'' set of the topologically different ones.

The latter, quantum-tunneling-related process of an abrupt change of
dominance was given the name of relocalization catastrophe in
\cite{arpot}. Several schematic $D=1$ illustrations of the
experimentally observable change of the subdomain were presented
there in detail. In our present upgraded, $D>1$ models the
mathematics of the construction (i.e., of the search for the
crossings of the locally dominant would-be ground states $E_0$)
remained the same.

What appeared to be new and different is the rich more-dimensional
structure of the domain ${\cal D}$ as well as of its subdomains
${\cal D}_\iota$ which are characterized by the specific, different
lists of the dominant (i.e., absolute) minima of $V(x,y,z)$.
From the quantum-world perspective,
a move between any two
parametric subdomains  ${\cal D}_A$ and ${\cal D}_B$
alters
the depths of the minima and
is accompanied by the (avoided) crossings
of the respective candidates $E_0^{(A)}$ and $E_0^{(B)}$
for the global ground state energy.

Naturally, the phenomenological wealth of the models should be kept
in balance with the intuitive transparency of the formulae.
If achieved,
the predictions of the abrupt
changes of the
probability density distributions
could really open a new branch of the theory,
especially if one manages to
localize the
quantum-catastrophic boundaries between the separate,
topology-protected subdomains in the space of parameters ${\cal D}$.
In parallel, this
might also prove deeply satisfactory from the
experimentalist's point of view.

\section{Summary\label{summary}}

In all of our present cusp ($N=1$)
and butterfly ($N=2$) models
in dimensions $D=2$ and $D=3$
we were able to specify
the admissible, well-behaved domains of
parameters
inside which the
qualitative
structure and behavior of the low-lying spectra
of bound-state energies and wave functions
was ``regular'', approximately separable and
fairly well approximated
by the closed, harmonic-oscillator-type formulae.
Naturally, whenever we tried to move closer to the
``prohibited''
boundaries of these domains,
we observed the loss of
reliability of our approximations.
This made our
basically non-numerical
polynomial-interaction
picture of the related
physical reality consistent.

We kept in mind that
due to the non-separability of our
partial differential Schr\"{o}dinger equations
an intuitive insight in some global features of the system
might be misleading.
Typically, for potentials admitting
the existence of multiple local minima it may be difficult to
localize the abrupt changes of the topological structure of
probability densities $\{|\psi_m(x,y)|\}$ or $\{|\psi_m(x,y,z)|\}$,
especially when they happen to be
caused by a minor variation of the coupling constants.
This was the key problem which we addressed.

We were inspired by the observation that in $D>1$ dimensions
and in the case of non-separable phenomenological potentials
a way of circumventing technical
obstacles is most often
sought in the restriction of attention to the
models which are
separable \cite{Winternitz}. In our present paper we
weakened such a pragmatic constraint.
We only assumed the form of separability
which emerges far from the origin, asymptotically
(cf. Eq.~(\ref{lyacur})).

For the sake of brevity we also
considered just the first few nontrivial
non-separable-interaction
scenarios. We assumed that $D=2$ and $D=3$ and
that $V \approx r^4$ or $V \approx r^6$
at the very large
distances from the origin, i.e., at $|r|\gg 1$.
We
admitted the general
polynomial forms of the potentials while requiring that
these functions
of coordinates were oscillatory.
In the low-energy regime such a
strategy was rewarded by the
not quite
expected
approximate separability
and approximative local
solvability of Schr\"{o}dinger equations.

Marginally, let us add that our choice of models led,
serendipitously,
also to a few
interesting mathematical benefits.
The first one
reflects our analyticity assumptions
and consists in
the user-friendly nature of description of the
shape of the potentials in question.
Another benefit
is connected with the
assumption of the locally pronounced nature of the individual
potential valleys and barriers.
As its byproduct we obtained
the models of artificial molecules
which proved tractable, in a fairly good approximation,
non-numerically.

Summarizing, after
the present transition from elementary $D=1$
to higher dimensions $D=2$ and $D=3$
we have to appreciate the survival of the transparency
of the resulting approximate non-numerical formulae.
This is certainly opening the way toward several
practical applications, especially in the brand new
context of quantum catastrophes.
In this context
our constructions of the
low-lying quantum bound states may be expected to
prove useful and instructive, especially due to their
phenomenologically promising
multi-quantum-dot nature.

\end{document}